\documentclass[runningheads]{llncs}
\usepackage{graphicx}
\usepackage{todonotes}
\begin{document}
\title{Deploying Crowdsourcing for Workflow Driven Business Process - a Brief Proposal}
\author{Rafał Masłyk\inst{1}\orcidID{0000-0003-1180-2159} \and
Kinga Skorupska\inst{1}\orcidID{0000-0002-9005-0348} \and
Piotr Gago \inst{1}\orcidID{0000-0001-7288-4210}\and
Marcin Niewiński\inst{1}\orcidID{0000-0002-6416-3541} \and
Barbara Karpowicz \inst{1}\orcidID{0000-0002-7478-7374}\and
Anna Jaskulska \inst{2}\orcidID{0000-0002-2539-3934} \and
Katarzyna Abramczuk\inst{3}\orcidID{0000-0003-2249-9888} \and
Wiesław~Kopeć\inst{1}\orcidID{0000-0001-9132-4171}} 
\authorrunning{Masłyk et al.}
\titlerunning{Deploying Crowdsourcing for Workflow Driven Business Process}
\institute{Polish-Japanese Academy of Information Technology \and
KOBO Association \and University of Warsaw, Faculty of Sociology}
\maketitle              % typeset the header of the contribution
\begin{abstract}
The main goal of this paper is to discuss how to integrate the possibilities of crowdsourcing platforms with systems supporting workflow to enable the engagement and interaction with business tasks of a wider group of people. Thus, this work is an attempt to expand the functional capabilities of typical business systems by allowing selected process tasks to be performed by unlimited human resources. Opening business tasks to crowdsourcing, within established Business Process Management Systems (BPMS) will improve the flexibility of company processes and allow for lower work-load and greater specialization among the staff employed on-site. The presented conceptual work is based on the current international standards in this field, promoted by Workflows Management Coalition. To this end, the functioning of business platforms was analysed and their functionality was presented visually, followed by a proposal and a discussion of how to implement crowdsourcing into workflow systems.

\end{abstract}
\section{Introduction}
The development of computer networks, the availability of devices such as personal computers and mobile communication offer great opportunities for cooperation between people who have never physically met. Thanks to this technology, they can make large and unique contributions to complex processes, for example, with crowdsourcing. \cite{howe2008crowdsourcing}

Business processes, often geographically dispersed and surrounded by many rules, may require technological support - through workflow systems. However, these systems, in a very strict manner, allow only certain user behaviors limited by their designed functionality. Lack of flexibility in terms of process changes during implementation, not allowing people or applications to perform tasks that are not foreseen in the model (and data), is a frequent criticism of this technology. Therefore, in this paper we explore how to expand the functional capabilities of typical business systems by allowing specific tasks to be performed by massive human resources, in this case with the use of crowdsourcing.

\section{An overview of methods and practice}
\subsection{Crowdsourcing platforms}
At this stage it is possible to present a certain scheme of operation of crowdsourcing platforms. First, the employer creates a campaign, according to the requirements of the crowdsourcing platform. From now on, the platform acts as a mediator in assigning tasks to the crowd or a human cloud. The client does not choose the worker for the specific task, but presents the planned tasks to the crowd of workers, who may or may not have to perform them. Users of the crowdsourcing platform who have successfully completed the registration phase can operate in two ways. They can be either employers or workers (human cloud), or both at the same time. According to data from the Microworkers.com platform, 90.52\% of all users act as workers, 3.59\% are people who engage only and exclusively as employers, while 5.89\% of all users are people acting on two fronts - as workers and employers simultaneously\cite{Franklin:2011:CAQ:1989323.1989331}\cite{5976179}. 

\subsection{Workflow management systems}
There are many systems on the market that allow for workflow management. Most very large companies develop such systems individually in order to have full control over the organization. They are generally called Workflow Management Systems\cite{Allen2001WorkflowAI}. Their structure and functionality define the standards developed by Workflow Management Coalition (WfMC) in the so-called Workflow Reference Model (WRM). Based on our preliminary research and previous work at this stage of we do not see systems allowing to resolve specific types of tasks with the help of crowdsourcing platforms or any other form of crowdsourcing \cite{hollingsworth_1995}. 

\section{Integrating Crowdsourcing Interaction into Business Process Management Platforms}
Here we elaborate on the problem of performing a crowdsourcing tasks with the use of workflow.

\subsection{Delegation of the task by the user to an external PCS}
Many Business Process Management Systems (BPMS) contain all the necessary tools to be able to design a workflow for all BPMS users. The question we pose is whether and how this functionality can be extended to include CS-type tasks.

The first solution is to use the existing forms of crowdsourcing to solve tasks. So let's assume that the indicated BPMS platform user is obliged to perform the task using crowdsourcing. We further assume that this task is suitable for execution by delegating its execution outside the organization, i.e. by crowdsourcing. For example, it could be a project of employee business cards.

Since this user does not have a mechanism enabling access to crowdsourcing (CS) platforms via his user panel in the BPMS system, the  task should be ordered individually, e.g. by means of a web browser on the CS platform. It should be emphasized here that the workflow type system monitors only two steps: start and finish. The user can pass this task on to an external company, some system or another user. The user is also obliged to interact with the workflow system in order to communicate the end of the task.

After the time allocated to this task has elapsed, the user verifies the results. For this purpose, he or she reviews the received graphic suggestions and selects the best ones. After paying for the accepted projects, the user downloads them. Then BPMS introduces the best results in its system and changes the status to closed. In this way the company received designs for its business cards. The task was done with CS but not with the use of the system and not based on the workflow.

It is obvious that there are no interactions between the workflow system and the CS platform indicated by the user. As a result, this method is a delegation of the task to an external CS platform.

The problem raised here, i.e. lack of communication between the BPMS system and the external CS platform, is quite important. Despite the generalisation of the principle of operation of CS platforms, building a system that creates a bridge in communication between BPMS systems and such platforms is not trivial. The lack of standards in this area, which would give guidelines for designing a possible communication platform based on workflows technology, does not allow us to propose a universal solution. Each CS platform has a different specification of the tasks, availability of contractors, forms of compensation for the task. There is no international body responsible for the standard approach to the construction and implementation of CS platforms.

\subsection{Proposed Workflow Systems Modifications}

\begin{figure}
    \centering\includegraphics[width=11cm]{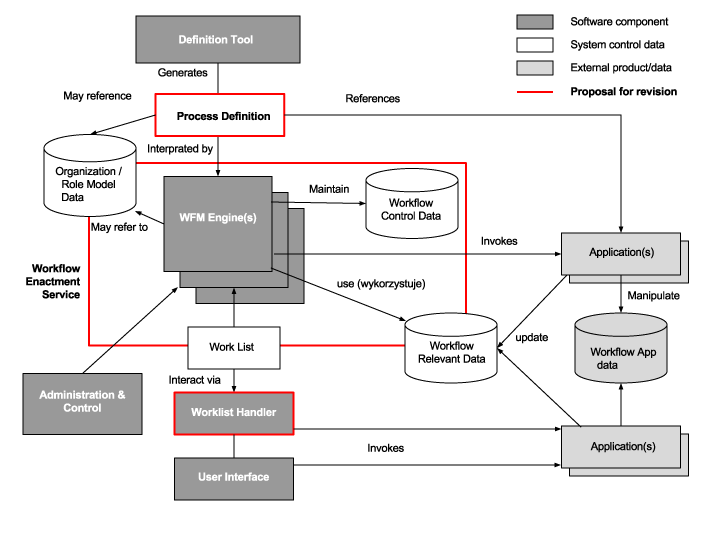}
    \caption{Proposed changes in the structure of the workflow system. Based on the WfMC workflow architecture.}
    \label{fig:added}
\end{figure}

Figure \ref{fig:added} shows the structure of workflow systems according to the Workflow Management Coalition, Reference Model, along with the indication of segments (objects marked in red) requiring modification in order to obtain additional functionality necessary to support cooperation with CS tasks.

The first change concerns Process definition. According to this standard, this component contains all the information about the structure, constraints, tasks, types, roles and other process defining data. This definition is the basis for the operation and feeding the necessary data of the most important component of this architecture, i.e. Workflow Enactment Service. It is composed of:
\begin{itemize}
\item	information on the conditions for starting and ending the process,
\item	activities and rules allowing for the continuation of the execution of the trial,
\item	the tasks the user can undertake, i.e. roles,
\item	references to applications that can be called,
\item	the definition of any WF data to which the system may refer.
\end{itemize}

In the case considered in this paper, the first change concerns the introduction of a definition of a new type of tasks [5] [6]. We will call them crowdsourcing tasks. As the name suggests, it is to be available to any external entity, organization or person as yet undefined in the user system.

The second change concerns the Task List Manager (Worklist Handler), we will continue to use the abbreviation WH. This component is responsible for managing the interaction between WF users and workflow enactment service. Its task is to take care of work progress within the whole process by interacting with the workflow enactment service (hereinafter referred to as WES) by means of a list of tasks (worklist) available for execution. In the case of some systems, this component may be more than just a system tool to inform about the available tasks to be performed. In the case of this work, WH must ensure that the crowdsurcing task is accessible to any external user and should remain accessible regardless of the number of people undertaking the task.

\subsubsection{How Worklist Handler (WH) Works}

Let's consider a simple example of a process definition, where a process consists of four tasks in the form of a sequential order.

\begin{figure}
    \centering\includegraphics[width=8cm]{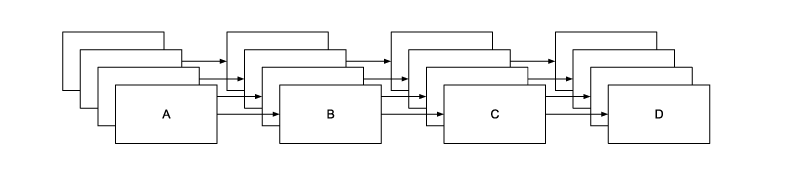}
    \caption{Process instances in WF Systems (Process Execution).}
    \label{fig:results2}
\end{figure}

As soon as task A becomes available for execution and is accepted by the user authorized to perform it, the system creates a copy of the entire process for the execution of this initiated instance of the process, as shown in Figure \ref{fig:results2}.
If the indicated task of the process is to be performed as a task of the CS type, as in all other cases, the initiation of the process instance is equivalent to the creation of a copy of the entire process. In addition, however, we have here potentially multiple execution of a single activity under execution of single process instance, for which standard WF systems are not prepared.

\begin{figure}
    \centering\includegraphics[width=8cm]{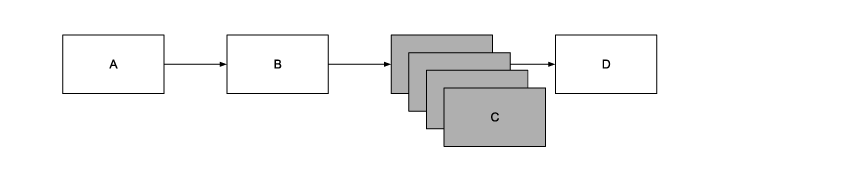}
    \caption{Multiple execution of a single activity under execution of single process instance.}
    \label{fig:results3}
\end{figure}

Figure \ref{fig:results3} illustrates such a case for a single process instance where the C task is of the CS type.
The multiple of asynchronous execution of task C results from the different time in which an indefinite number of people can start performing this task. In the extreme case, the task did not interest anyone and then it will not be done at all. It is rather obvious that it is impossible to predict how much copying is necessary for the management of this task. The system must generate another copy together with the registered needs, i.e. the next performers of this task who begin activities in this area.

\begin{figure}
    \centering\includegraphics[width=11cm]{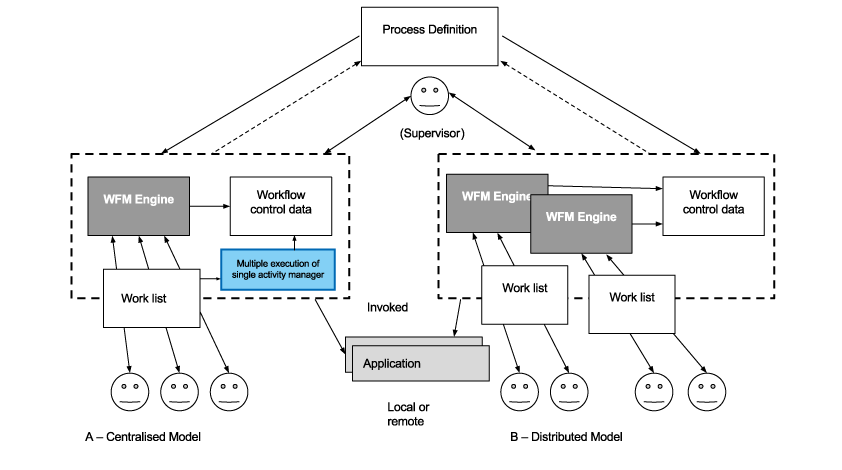}
    \caption{Standard workflow enactment service area (WES) with the new MESAM component.}
    \label{fig:results4}
\end{figure}

Now we have to look at the next BPMS component. This is the Workflow Enactment Service (WES) shown in Figure \ref{fig:results4}. To allow a CS task to be performed by a standard WF system, it is necessary to modify the WES.

The workflow enactment service would need a new element to manage \textbf{multiple executions of a single activity.}, from now on called Multiple Executions of a Single Activity Manager (MESAM). This extension would be responsible for the asynchronous management of multiple existence of a single instance. Adding this as a separate element is intended to showcase that not every task should be performed using CS, therefore it is not always necessary to use this extension. 

In the process of creating subsequent executions of a single activity within the execution of one instance of the process, the machines of states will also be duplicated. However, the process remains in the active state during the designated time, when new instances can be initiated. It should be emphasized here that initiating the execution of successive instances of this task is asynchronous and devoid of communication between the users of the service created in this way. Each of these machines is responsible for the current state of a single copy of a given activity within the framework of a crowdsourcing task. Let us remind that the limitation of the CS type task is to be the duration, i.e. the openness of the task for potential contractors, therefore, a certain number of machines will be created, which are in different states of activity. Their number will depend on how many system users have undertaken this task. After a certain amount of time (formulated / set in the conditions of this task), this task will be completed at the process level.

\begin{figure}
    \centering\includegraphics[width=8cm]{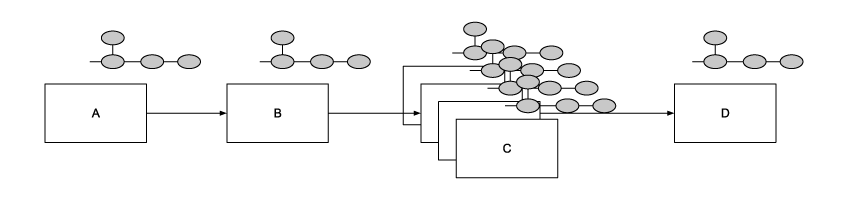}
    \caption{Multiplication of the machinery of the states of activity.}
    \label{fig:results6}
\end{figure}

At this point, the proposed component - MESAM, will face the task of completing all instances of the activity under consideration.
To solve the problem, the proposal is to force the termination of all unfinished tasks (in the Active state) of this activity.

In order for the proposed concept to be implemented in practice, it is necessary to introduce a new type of task, a task of the CS type. Such a change in the meta-data definition of the process enforces another modification related to the presentation of such a task, i.e. an important part of the architecture of the standard workflow system - worklist hendler (WH).

It seems natural that with the new type of CS task, the WH component should store into a dedicated form of operation. Our proposal is as follows: WH would then serve as an interface for undefined users\cite{inproceedingsorl} \cite{inproceedingsoorla} in the workflow system to redefine a schema for retrieving information about users, such as identification data, id assignment, contact data, and other data. A task of the CS type would still be available within the framework of the defined task, the time for its execution. This data must be securely stored in the system for a longer, specified period of time. 

\begin{figure}
    \centering\includegraphics[width=11cm]{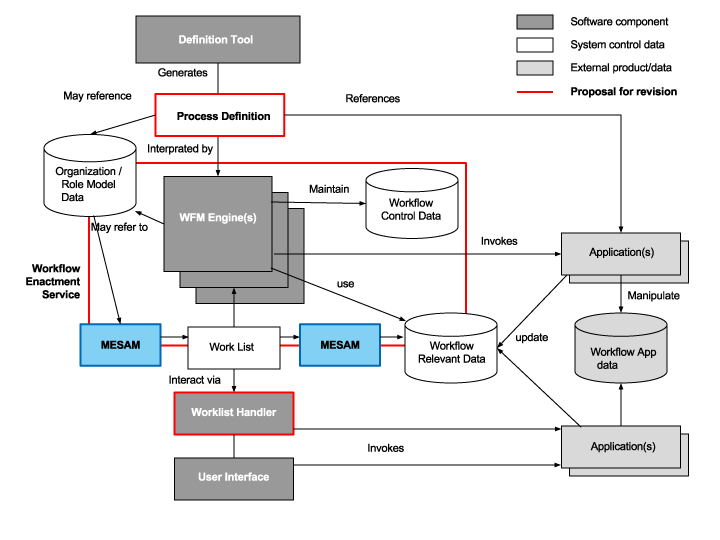}
    \caption{Proposed changes in the workflow structure.}
    \label{fig:results7}
\end{figure}

Thanks to this, during the system's operation, the expectations concerning data consistency and continuity at each stage of the WF system's operation are met. Naturally, this requires the expansion of the database structure, its properties and limitations, and the expansion of the above mentioned interface. Figure \ref{fig:results7} shows a proposal to place new components (marked in blue) in the workflow system structure.

To sum up, the number of necessary modifications to BPMS operational work is significant. Each aspect mentioned requires precise specifications in order to be integrated into existing standards.

\section{Conclusions}
In this paper we discussed how to integrate the possibilities of crowdsourcing platforms with systems supporting workflow to enable the engagement and interaction with business tasks of a wider group of people. Lack of flexibility in terms of process changes during implementation, allowing people or applications to perform tasks that are not foreseen in the model (and data), is a frequent criticism of this technology. Thus, this work is an attempt to expand the functional capabilities of typical business systems by allowing selected process tasks to be performed by unlimited human resources.
Our proposal is to extend the current standards with the necessary elements in order to achieve the goal of allowing the task to be carried out as an element of the process, not by designated persons from the organization, but by crowdsourcing. Inclusion of a new type of task requires a number of changes in the workflow system modules. These are elements of the definition of a task in the process, which requires a change in the meta-data of the system and the adaptation of the engine of the system to the execution of this type of procedure. This change entails changing an important module such as Work List Handler, a module responsible for organizing orders in accordance with the process definition. 

Opening business tasks to crowdsourcing, within established Business Process Management Systems (BPMS) will improve the flexibility of company processes and may allow for lower work-load and greater specialization among the staff employed on-site. When implemented well it may also mean an increase in overall task completion quality and a potential decrease in the monetary cost of the task. However, there will be some costs associated with the application of the proposed process related to implementation and marketing, ongoing management and monitoring, including quality assurance. While these are important considerations for all companies, the change for external contractors, such as crowd workers, is overall positive, as it would mean more diverse work opportunities facilitating competence-building and increased specialization.

Our further work will focus on expanding this model and testing its feasibility in the business context and with different types of crowdsourcing tasks, either completed by human resources or in a hybrid manner (using, for example, machine translation verified by human crowd workers, as postulated in our previous work \cite{skorupska_warpechowski_nielek_kopec_1970} \cite{skorupska19}). Overall, we believe there is massive potential in extending the use of crowdsourcing in business contexts, and especially within established Business Process Management Systems (BPMS).

\section{Acknowledgments}
The presented conceptual work is based on the current international standards in this field, promoted by Workflows Management Coalition and was made possible by the XR Lab at PJAIT, part of HASE research initiative (Human Aspects in Science and Engineering) of human-computer interaction labs.

\bibliographystyle{splncs04}
\bibliography{bibliography}
\end{document}